\documentclass[conference]{IEEEtran}
\ifx\pdfoutput\undefined
\usepackage{graphicx}
\else
\usepackage[pdftex]{graphicx}
\fi

\ifCLASSINFOpdf
\else
\fi
\usepackage[latin2]{inputenc}

\begin{document}
%
\title{High-Resolution, Multi-Channel FPGA-Based Time-to-Digital Converter}


\author{\authorblockN{Bal\'azs J\'akli\authorrefmark{1},
\'Ad\'am R\'ak\authorrefmark{2} and
Gy\"orgy Cserey\authorrefmark{1}}
\authorblockA{\small\authorrefmark{1}Faculty of Information Technology and Bionics, P\'azm\'any P\'eter Catholic University, Budapest, Hungary\\
Email: cserey.gyorgy@itk.ppke.hu}
\authorblockA{\small\authorrefmark{2}MedioTrade Ltd., Budapest, Hungary}}


%


\maketitle

\begin{abstract}
In this paper we present a novel high-resolution multi-channel FPGA-based time-to-digital converter (TDC). We designed and implemented a complex electronic circuit on the FPGA, whose overall accuracy is several orders of magnitude greater than the accuracy of the FPGA used in digital mode. Our sensor device contains simple circuit elements that are cheap and easily accessible (Xilinx Spartan 3 and Spartan 6). Using our design, many channels (80-100 channels) can be implemented on a larger FPGA. The prototype of our TDC has been implemented and functionally verified by experiments and measurements. By a certified pulse generator 20 ps precision has been measured over the range of 3 ns. Using more precise clock signal this range may be extended. The achieved resolution is 5 ps.  Its resolution, channel number and range can be configured dynamically, which makes it suitable for effective use in industrial purposes.
\end{abstract}


%
\IEEEpeerreviewmaketitle

\section{Introduction}

Integrated circuits which can measure time-intervals with high precision are called time-to-digital converters (TDC). 
The majority of time measurement tools on the market primarily use only a few channels (1-8 channels). A channel is characterized by two inputs, it measures the time delay between their signals. 
TDC has applications in various fields of industrial engineering. It is used for LIDAR~\cite{nissinen2009chip}, medical imaging applications~\cite{yousif2007fine} and time-of-flight mass spectrometry as well as it is used in logic analyzers, where extremely high resolution and accuracy are needed. 
Multi-channel TDC can be specifically apply to determine the 3D spatial localization of events with cm precision in a particle physics detector. 
Until the recent years, most TDC designs have been developed and implemented for specific applications, but the latest trends shows that configurable circuit implementations can be successful using field programmable gate arrays (FPGA)[6-13]. The advantages of the FPGA are lower development time and lower cost in small series. 
Many approaches have been tried to use the opportunities of the FPGA to measure time [6-13]. One of the latest FPGA-based TDCs is documented in paper~\cite{daigneault2011high} that is close in precision to our solution, but they used the Virtex II Pro FPGA, which is more expensive, and it is not suitable for the realization of multi-channel TDCs. \cite{wu2007adc} and \cite{bocci2007multichannel} are presenting multi-channel TDCs (about 96 channels), but their resolution has the order of magnitude of 1 ns.

The paper is organized as follows: Section II describes our motivation. Section III presents the proposed architecture. In Section IV experimental results demonstrate the features of our TDC design. Finally, the conclusions and future work are summarized in Section V. 

\section{Motivation}

The measurement of time is essential in electrical measurements, practicing engineers use these techniques in many cases. An analog-digital conversion is performed by a TDC. The analog input value is a relative time difference of two events between each other, such as the delay between two signals. The output is a value in a discrete range, which is the approximated result of the converted input value by quantization. The unit of the conversion, the resolution of the TDC is the shortest time which makes changes, either increasing of decreasing of the output value. Currently, this magnitude is in pico-second range.

The number of TDSs have reached a marketable level by the very large scale integration of circuits, but the vast majority of commercially available devices have only a few channels (channels 1-8). A channel is characterized by two inputs, the relative temporal delay is measured between the inputs. These sensors are implemented on ASICs (Application Specific Integrated Circuits). Multi-channel TDCs are difficult and expensive, but there are tasks where they can be used effectively: for example, using TDCs in particle detectors, 3D spatial localization of events can be determined with precision in the order of cm, and also their usage can be necessary in other particle physics experiments. Nowadays these particle detectors are very expensive and have only a relatively small accuracy. A cost-effective, high precision, multi-channel TDC would be suitable to fulfill a currently empty market segment. These TDCs would be used to complete measurements, which could not have been done by physicists previously, as well as in multi-channel logic analyzer devices, where high resolution and high accuracy is required.

\section{Proposed architecture}

One of the fundamental innovations in our approach is that rather than only compare the two input signals in a channel, we have included the many calibrated reference signals as well in our comparison, and thus, we have been able to achieve high precision and low-noise in our measurements. After comparing the calibrated signals to the input signals, the resulting absolute values render a measurement range that is theoretically infinite and practically very large. In the event that we have multi-inputs, they are simultaneously compared to the reference signals. Another innovation is the design and measurement method of the internal structure of the FPGA, which provides a stochastic bit-stream based time-to-digital converter (TDC). We can reconfigure our system dynamically by changing the channel number or measurement accuracy of each channel even during operation. The expected achievable accuracy of the system is ~1ps.

\begin{figure*}[ht]
\begin{center}
\includegraphics[angle=0, width=195mm]{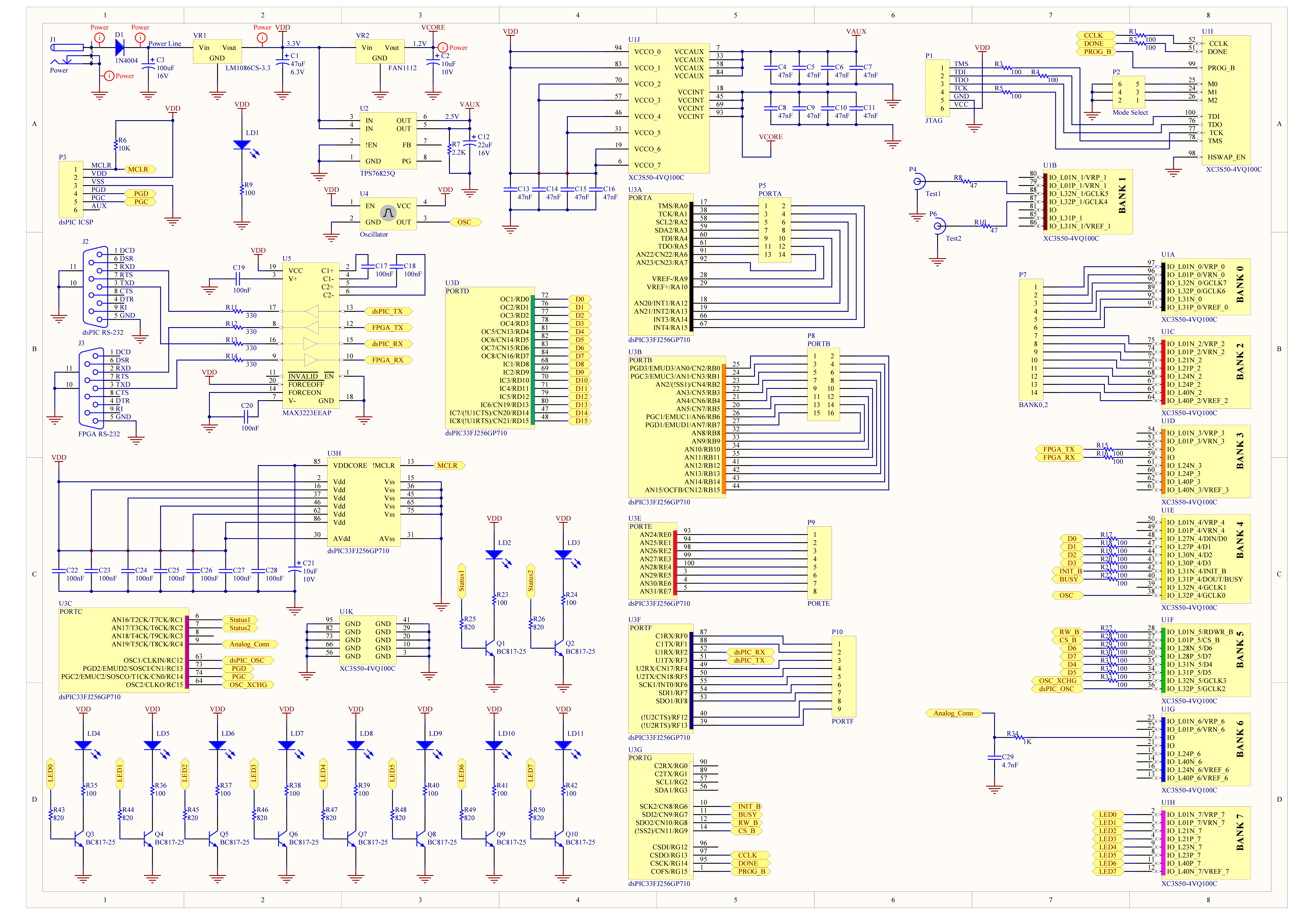}
\caption{
Schematic of the prototype board.}
\label{figure:prototype_board}
\end{center}
\end{figure*}

\begin{figure}
\centering
\includegraphics[width=3.4in]{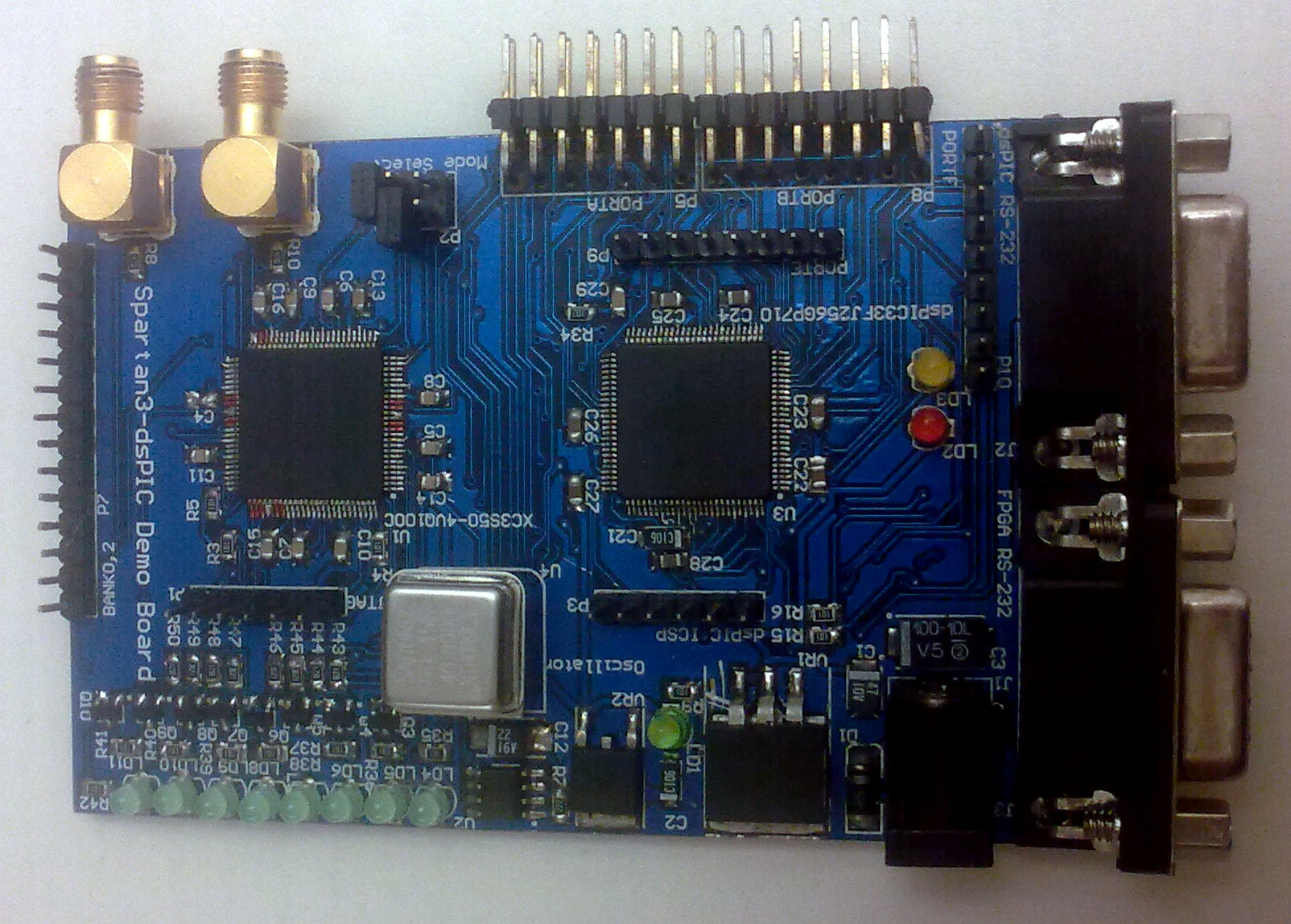}
\caption{Implemented prototype.}
\label{fig_sim3}
\end{figure}

We designed and implemented an experimental prototype of our device (see Fig.~\ref{fig_sim3}). The schematic can be seen on Fig.~\ref{figure:prototype_board}. 
Besides the Xilinx Spartan 3 FPGA chip, we placed a dsPic to do the control and communication to the FPGA chip. This solution gives the opportunity to re-configurate the channel number and the size of the actual measurement range of our system.

\section{Experimental Results}

\begin{figure}
\centering
\includegraphics[width=3.4in]{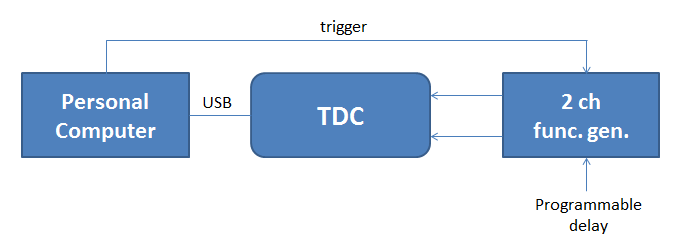}
\caption{
We measured the pulse delay between two signals generated by a two channel function generator (Type: Agilent 81130A) A personal computer triggered the signals, then read out the output values of the TDC conversion via USB interface.
}
\label{fig_sim1}
\end{figure}

We measured the pulse delay between two signals generated by a two channel function generator (Type: Agilent 81130A) A personal computer triggered the signals, then read out the output values of the TDC conversion via USB interface (see Fig.~\ref{fig_sim1}). 

\begin{figure}
\centering
\includegraphics[width=3.4in]{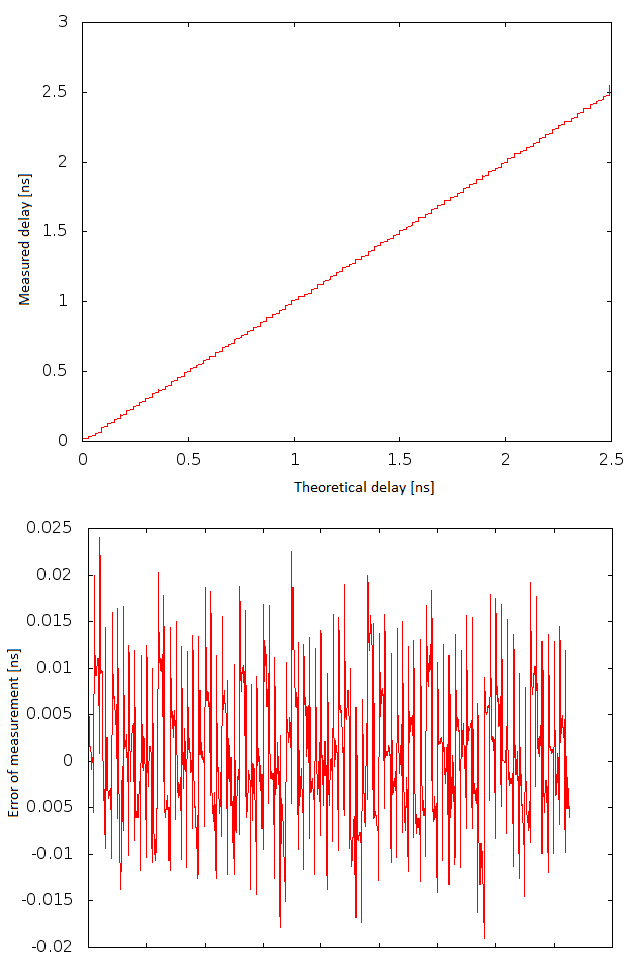}
\caption{
On the upper side we see the output of the TDC conversion depending on the delay set on the function generator. At the bottom side, we see the error between the theoretical and the measured delays. As we can see, the accuracy of the measurement is about 20 pico-secundums, which is the accuracy of the function generator. It is assumed that our method is capable of even more accurate results.
}
\label{fig_sim2}
\end{figure}

On the upper side of Fig.~\ref{fig_sim2} we see the output of the TDC conversion depending on the delay set on the function generator. At the bottom side of Fig.~\ref{fig_sim2} , we see the error between the theoretical and the measured delays. As we can see, the accuracy of the measurement is about 20 pico-secundums, which is the accuracy of the function generator. It is assumed that our method is capable of even more accurate results.

\begin{figure}
\centering
\includegraphics[width=3.4in]{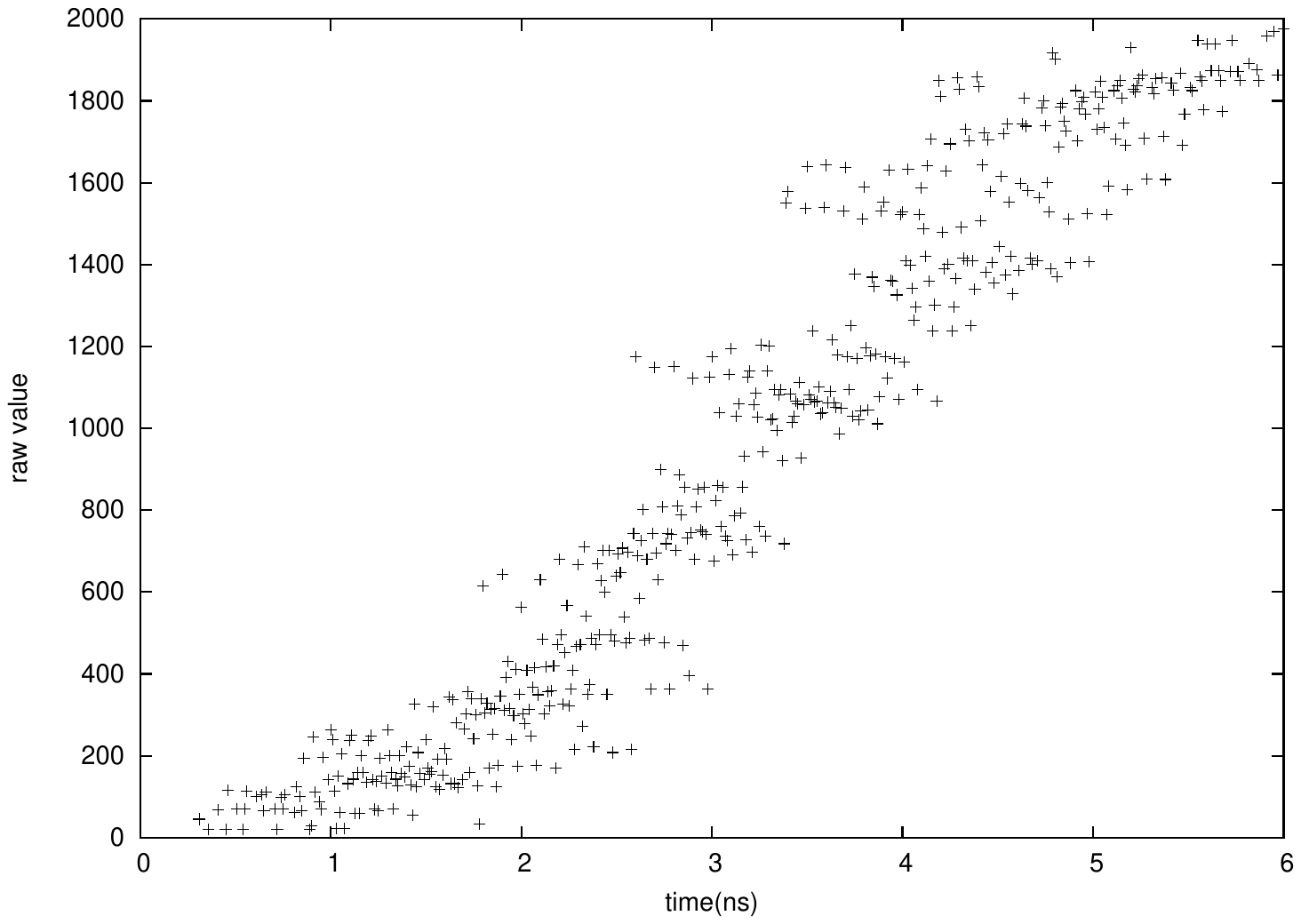}
\caption{
Measured raw uncalibrated data from our TDC is depicted on the figure versus the input time difference of the impulses.
}
\label{fig_meres}
\end{figure}

The experimental prototypes of our device have been implemented and its functionality is confirmed by experiments and measurements (see Fig.~\ref{fig_meres}). We have captured the raw stochastic bitstream output of our TDC. In our first results we did not use any calibration on the bitstream, instead we have simply summed the bits as integers. This summation decreased the stochastic noise enough to allow us to see the working time measurements as seen in Fig.~\ref{fig_meres}. We have also tested the shifting of our measurements range and it worked exactly as precise as the clock source we have used as a reference source for shifting by whole clock cycles therefore extending the range to practically infinity (the precision is limited only by the frequency stability and the jitter of the clock source). Our latest prototype was applied on the Xilinx Spartan 6 FPGAs. The parameters of our innovative sensor can be configured flexibly. 

\begin{figure}
\centering
\includegraphics[width=3.4in]{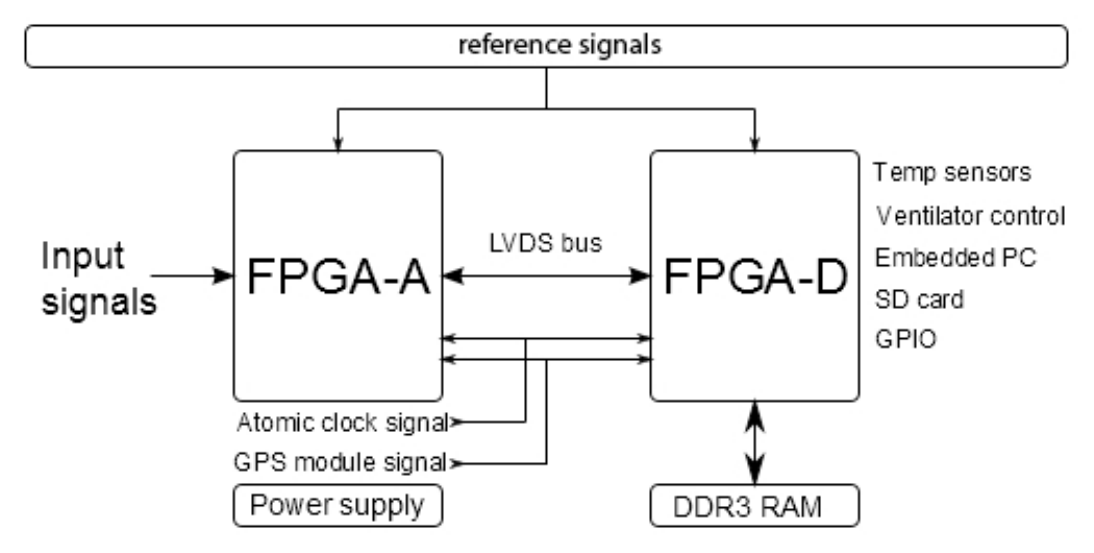}
\caption{
FPGA-A does the time-to-digital conversion. FPGA-D is the control FPGA responsible for the temperature-stability, control processes and measurements of the reference signals and their parameters. 
}
\label{fig_board}
\end{figure}

\section{Conclusions and future work}

The experimental prototype of our device has been implemented (see Fig.~\ref{fig_sim3}) and its functionality is confirmed by experiments and measurements (see Fig.~\ref{fig_sim1}). We measured 20 ps accuracy rate by a certified pulse generator (20 ps, where the range is 3 ns, see Fig.~\ref{fig_sim2}). The range of the measurement may be extended by using a more accurate clock signal. Our prototypes were applied on the smallest Xilinx Spartan 3 and Spartan 6 FPGAs. Using a greater FPGA, up to 100 high-precision channels can be implemented. The parameters of our innovative sensor can be configured flexible. Using active cooling (Peltier-element) the range of the sensor system can be extended, while operation range becomes more stable.

\section*{Acknowledgment}
The invaluable help of the multidisciplinary doctoral school at the Faculty of Information Technology of the P\'azm\'any P\'eter Catholic University is gratefully acknowledged. The authors acknowledge the support of the Hungarian Government through grant no. VEKOP-2.1.7-15-2016-00550. 


%
%



%
%
%

\nocite{*}
\bibliography{refs} 
\bibliographystyle{IEEEtran}

\end{document}